\documentclass[referee,sn-nature,oneside]{sn-jnl}

\usepackage{graphicx}%
\usepackage{adjustbox}
\usepackage{multirow}%
\usepackage{amsmath,amssymb,amsfonts}%
\usepackage{amsthm}%
\usepackage{mathrsfs}%
\usepackage{xcolor}%
\usepackage{textcomp}%
\usepackage{manyfoot}%
\usepackage{booktabs}%
\usepackage{algorithm}%
\usepackage{algorithmicx}%
\usepackage{algpseudocode}%
\usepackage{listings}%
\usepackage{lipsum}%
\usepackage{gensymb}%
\usepackage{float}
\usepackage[left]{lineno} %
\usepackage{changepage} %
\usepackage[T1]{fontenc} 
\usepackage{anyfontsize} 
\usepackage{caption}

\setcitestyle{numbers,square,comma,sort&compress}
\begin{document}

\title[Article Title]{Deterministic nanofabrication for engineering nanowire quantum dot devices}


\author*[1,2]{\fnm{Tarun} \sur{Patel}}\email{t24patel@uwaterloo.ca}
\equalcont{These authors contributed equally to this work.}

\author[1,2]{\fnm{Matteo} \sur{Pennacchietti}}
\equalcont{These authors contributed equally to this work.}

\author[3]{\fnm{Greg} \sur{Holloway}}

\author[1,2]{\fnm{Stephen} \sur{R. Harrigan}}

\author[1,2]{\fnm{Sayan} \sur{Gangopadhyay}}

\author[1,2]{\fnm{Anthony} \sur{Drouin}}

\author[5,6]{\fnm{Megha} \sur{Jain}}

\author[5,6]{\fnm{Dan} \sur{Dalacu}}

\author[5]{\fnm{Philip} \sur{J. Poole}}

\author[1,2]{\fnm{Sasan} \sur{Vosoogh-Grayli}}

\author*[1,2,4]{\fnm{Michael} \sur{E. Reimer}}

\affil[1]{\orgdiv{Department of Electrical and Computer Engineering}, \orgname{University of Waterloo}, \orgaddress{ \city{Waterloo}, \postcode{N2L 3G1}, \state{Ontario}, \country{Canada}}}

\affil[2]{\orgdiv{Institute for Quantum Computing}, \orgname{University of Waterloo}, \orgaddress{ \city{Waterloo}, \postcode{N2L 3G1}, \state{Ontario}, \country{Canada}}}

\affil[3]{\orgdiv{Quantum-Nano Fabrication and Characterization Facility}, \orgname{University of Waterloo}, \orgaddress{ \city{Waterloo}, \postcode{N2L 3G1}, \state{Ontario}, \country{Canada}}}

\affil[4]{\orgdiv{Department of Physics and Astronomy}, \orgname{University of Waterloo}, \orgaddress{ \city{Waterloo}, \postcode{N2L 3G1}, \state{Ontario}, \country{Canada}}}

\affil[5]{\orgname{National Research Council Canada}, \orgaddress{, \city{Ottawa}, \postcode{K1A 0R6}, \state{Ontario}, \country{Canada}}}

\affil[6]{\orgdiv{Department of Physics, Engineering Physics and Astronomy}, \orgname{Queen’s University}, \orgaddress{ \city{Kingston}, \postcode{K7L 3N6}, \state{Ontario}, \country{Canada}}}

\abstract{
Semiconductor quantum dots (QDs) are a leading platform for realising bright, wavelength-tunable sources of single and entangled photon pairs for photonic quantum technologies. Site-selected nanowire quantum dots (NWQDs) are a promising platform for fabricating such photonic devices in a scalable manner. However, implementing additional structures around the photonic nanowire while maintaining its vertical growth geometry has remained a challenge. In this work, we develop a deterministic pick-and-place technique to conduct a vertical-to-vertical transfer of NWQDs from the growth substrate to arbitrary templates. Using this transfer technique, we enhance the photon extraction efficiency to 75\,\% by implementing a bottom gold mirror and tune the emission wavelength by 3.6\,GHz via implementing electrostatic gates around the QD. Importantly, we measure low-multiphoton probability (g\textsuperscript{2}(0)=0.002) and high indistinguishability ({\normalfont >80\%} for $\pm$ 100\,ps) of the QD emission after the transfer process, yielding high-quality devices. These results demonstrate the repeatability and versatility of the developed transfer technique, which is an enabling step towards scalable single and entangled photon sources.

}

\maketitle
\section*{Introduction}

The ability to engineer the properties of a quantum emitter coupled to a nanophotonic structure is critical for applications in quantum computing \cite{maring2024versatile}, quantum networking \cite{laneve2025quantum} and distributed quantum sensing \cite{stas2026entanglement, gottesman2012longer}.
Epitaxially grown semiconductor quantum dots (QDs) have emerged as a leading platform for engineering quantum light sources with low multi-photon emission, near-unity quantum efficiency, and high indistinguishability. Nanophotonic structures such as micro-cavities \cite{wang2019towards}, hybrid circular Bragg gratings \cite{wang2019demand} and waveguides \cite{meng2024deterministic} have been employed to obtain high photon collection efficiencies from QDs for on-demand single-photon, entangled photon pair or photonic cluster state generation. 
Current efforts have also aimed at storing quantum information in the nuclear spin ensemble of the QDs \cite{appel2025many}, extending their functionality into multi-qubit quantum nodes.     

Despite this progress, there are still outstanding challenges for at-scale applications that require deterministic fabrication and control of multiple single-QD devices. The main challenges faced by epitaxially grown QDs have been obtaining high coupling efficiency between the nanophotonic structure and the QD while simultaneously having the ability to tune individual emitters to the same emission wavelength in a scalable manner \cite{Rickert2025-wx, Albrechtsen2026-no}. Nanowire quantum dots (NWQDs) grown via selective-area (SA) vapour–liquid–solid (VLS) epitaxy have a single InAsP QD positioned strictly along the axis of an InP photonic nanowire. The SA-VLS growth can thus be leveraged for deterministic positioning of NWQDs with near-unity yield \cite{Laferriere2022-aa}. The diameter of the photonic nanowire can be engineered to support a single guided electromagnetic mode to ensure high coupling efficiency with $\beta$-factors approaching 0.95 \cite{Haffouz2018-rd}. Additionally, the top of the waveguide is tapered at angles below $2^\circ$ to adiabatically couple the guided electromagnetic mode from the nanowire waveguide into a free-space Gaussian beam with near-unity transmission and coupling efficiency into a single-mode fiber (SMF) of over 90\%  \cite{Bulgarini2014-ma}. The growth parameters enable accurate ($\pm4\,$nm) \cite{Laferriere2021-kc} control over the QD s-shell emission wavelength from near-IR (870\,nm) to telecom c-band (1550\,nm) by tuning the QD size and composition. Furthermore, the high confinement symmetry of the QDs ensure that they have intrinsically low exciton fine-structure splitting (FSS), which is essential for entangled photon pair generation \cite{Versteegh2014-sd}. Together, these attributes make SA-VLS grown NWQDs a promising platform to realize practical devices in a scalable manner.   

However, it has been challenging to fabricate additional structures around the NWQD while maintaining its upright geometry.    
Specifically, implementing a bottom mirror can increase the photon extraction efficiency to near-unity from the as-grown limit of $50\%$, while fabricating electrostatic gates around the QD would enable tuning of the emission wavelength.           
In this work, we develop a pick-and-place nanowire transfer technique via nano-gluing to deterministically transfer NWQDs from their growth substrate onto arbitrary, pre-fabricated templates while maintaining the vertical geometry. 
Further, we demonstrate the utility and versatility of this technique by transferring the nanowires onto two different pre-fabricated templates: (1) a NWQD on a gold mirror with a $10\,\mathrm{nm}$ SiO$_{2}$ spacer to implement a weak nanowire cavity and increase the photon collection efficiency; and (2) a NWQD between pre-fabricated electrodes for tuning the QD emission wavelength. 
The nanowire transfer onto a gold mirror increased the photon extraction efficiency to $75\%$, and introduced a weak Purcell enhancement of $1.58$.
Using the electrode-based device, the QD emission wavelength is shifted by over $3\,$GHz.  
In both cases, the optical spectrum and quality of single-photon generation remains unaltered, demonstrating that the QD is preserved during the transfer.
This nanowire transfer technique opens up the possibility to repeatably fabricate NWQD devices that can eventually combine high collection efficiencies with wavelength tuning and QD charge state control.

\section*{Vertical Nanowire Transfer Method}

The NWQD transfer is performed in a commercial dual-beam scanning electron microscope/focused ion beam (SEM/FIB) system equipped with a gas-injection system (GIS). First, a tungsten tip is sharpened down to a few hundreds of nanometres using FIB milling to match the diameter of the InP nanowire. The prepared tip is controlled using a nano-positioner stage that is used to detach the nanowire from the growth substrate and perform a deterministic pick-and-place procedure.

\begin{figure}[t!]
    \centering
    \includegraphics[width=1\linewidth]{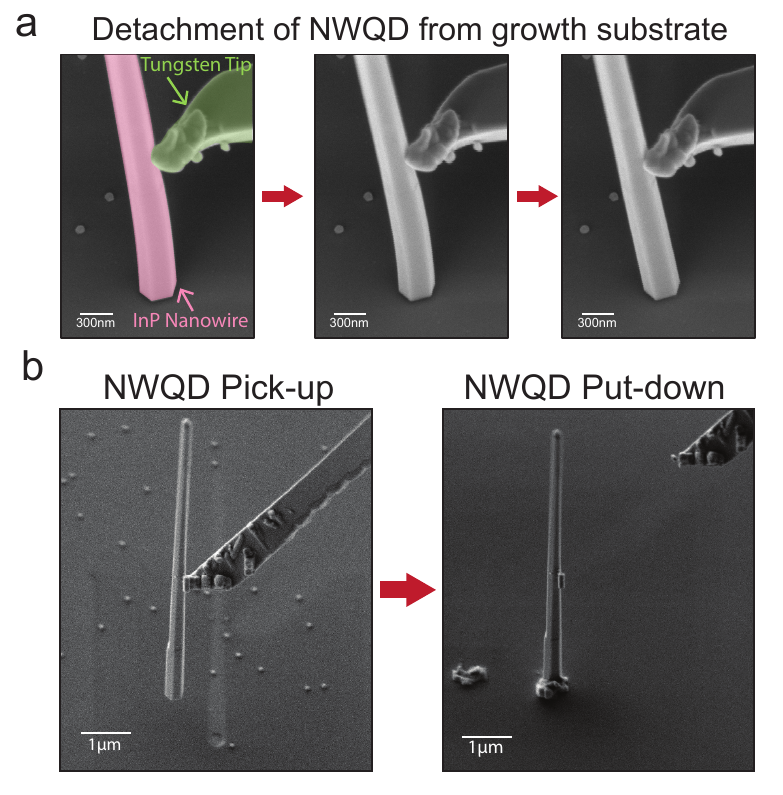}
    \caption{Visualizing deterministic pick-and-place transfer of a NWQD in a vertical-to-vertical geometry. (a) SEM images of the NWQD detachment from the growth substrate. From left to right, a tungsten tip (green) is slowly pushed against the InP nanowire (pink). The elimination of bending and the release of built-up elastic energy in the nanowire confirm its detachment from the growth substrate. (b) The detached nanowire is glued using focused electron beam induced deposition (EBID) of SiO$_x$ to the tungsten tip and picked-up (left panel). The nanowire is secured to the target substrate by additional SiO$_x$ deposition at the base and the tungsten tip is detached by pulling away from the nanowire (right panel).}
    \label{fig:figure1}
\end{figure}

The NWQD detachment process is shown in Figure \ref{fig:figure1}a. For a typical wire with a height of $10\,\mu $m, the probe is positioned approximately $1.5\,\mu$m above the growth substrate and comes in contact with the nanowire from one side, as shown in the left panel of Figure \ref{fig:figure1}a. The probe is then used to gently push the nanowire (center panel of Figure \ref{fig:figure1}a), causing it to bend and accumulate elastic strain energy. As the applied force increases, the nanowire eventually breaks at its base and detaches from the substrate. The detachment is evidenced by elimination of bending at the nanowire base, shown in the right panel of Figure \ref{fig:figure1}a. Once the break is confirmed, the tungsten probe is pulled away from the nanowire leaving it in the upright geometry. Since during its growth the nanowire overgrows the 100\,nm diameter oxide hole on the growth substrate, the diameter of the nanowire undergoes a drastic change at it's base providing a point of stress enhancement \cite{dalacu2012ultraclean}. Thus, the nanowire cleaves smoothly along a lattice plane as shown in SI Figure \ref{si_fig:nwqds_flat_cleave}. Minimizing the accumulated elastic energy is critical for maintaining contact between the nanowire and the tungsten probe throughout the detachment process. This is accomplished by carefully selecting the height along the nanowire at which the tungsten probe applies the force.  

After detaching the nanowire from the growth substrate it is attached to the tungsten probe using focused, nanoscale electron-beam-induced depositionp (EBID). The EBID involves injection of a precursor gas close to the sample, which then reacts with the focused electron beam allowing for controlled deposition of select materials at the nanometer tscale \cite{randolph2006focused}. For NW transfers shown in this manuscript, we use siloxane as the precursor gas to deposit SiO$_{x}$ as the nano-glue material. The nanowire is then glued to the tungsten tip at a height of 3 $\mu$m above the growth substrate, well above the QD height of 1.5 $\mu$m, thereby preventing any deformation or strain in the vicinity of the QD. As shown in the left panel of Figure 1b, the nanowire can now be picked up from the growth substrate using the aforementioned pickup process. Care is taken to minimize the amount of deposited material to glue the NW to the tungsten probe, as it is sacrificial. With the nanowire securely attached to the probe, the probe is retracted and the growth substrate is swapped out with an arbitrary target template. The nanowire is then placed at the relevant position on the pre-fabricated template, maintaining its upright geometry. Finally, the bottom of the nanowire is glued strongly to the target substrate using a larger deposition such that the initial bond between the tungsten probe and the nanowire can be broken by a pulling-away motion. A video recording of the transfer process is supplied with the manuscript. An example of this `put-down' step is shown in the right panel of Figure 1b. The EBID-assisted pick-and-place transfer process allows for the first time to deterministically transfer NWs in a vertical-to-vertical geometry. This represents a significant advancement over prior nanowire transfer techniques, which relied exclusively on van-der-Waals forces \cite{Zadeh2016-ft, Mnaymneh2020-lv} for `pick-up' and `put-down' and were limited to a lying-down geometry. The advantage of using a deterministic attachment process can also be extended to the lying-down geometry, highlighting the versatility of EBID. We now focus on two salient results obtained by implementing the NWQD transfer process that we have developed.

\bmhead{NWQD on a Au/SiO$_2$ Mirror}\hfill

For the first demonstration, we transferred eight NWQDs onto a gold substrate with an ultra-thin ($10\,\mathrm{nm}$) SiO$_2$ coating (see Methods). See SI Figure \ref{si_fig:nwqds_on_mirror} for an optical microscope image of the device. The motivation for such a device is to increase the collection efficiency of the QD emission to $\sim97\%$ \cite{claudon2010highly}. For NWQDs on the growth substrate the QD emission couples equally to the upward and downward propagating nanowire waveguide modes limiting the maximum collection efficiency from the top to $50\%$. By implementing a modal reflection at the bottom facet using the Au/SiO$_2$ substrate we can direct almost all of the QD emission in the upward direction via the adiabatic taper into a free space Gaussian mode.

\begin{figure}[h!]
    \centering
    \begin{adjustbox}{center,width=0.50\linewidth}
        \includegraphics{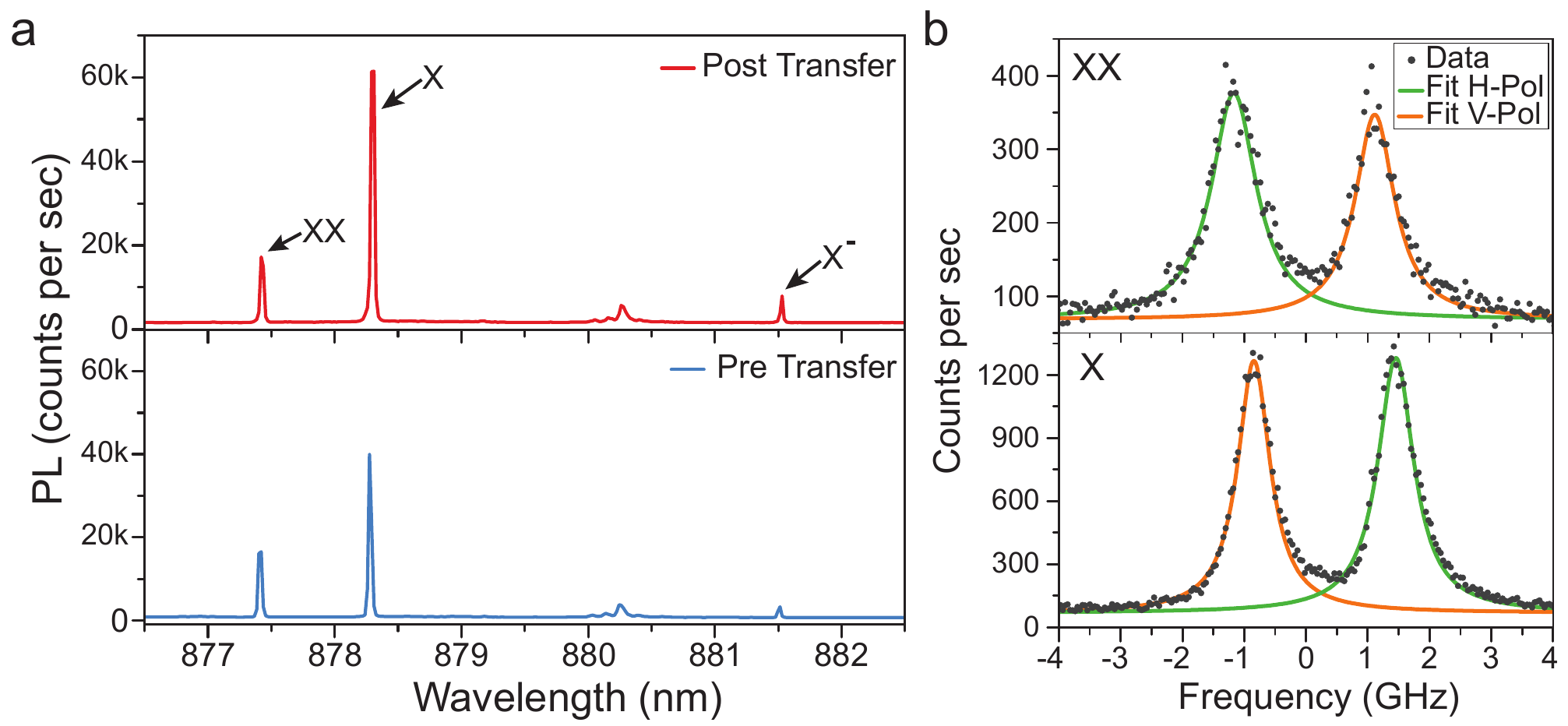}
    \end{adjustbox}
    \caption{(a) Photoluminescence (PL) spectra of NWQD5 before and after transfer on a Au/SiO$_2$ substrate under pulsed above-band excitation. The three prominent emission lines corresponding to the biexciton (XX), exciton (X) and negative trion (X$^-$) decay are labeled. (b) High resolution PL spectrum of the XX (top) and X (bottom) emission lines of the transferred NWQD5 measured using a scanning Fabry-Perot etalon. Two polarized peaks are observed due to the fine structure splitting (FSS) of the QD. Individual peaks are fit to a Lorentzian linewidth of $0.71 \pm 0.01\,$GHz (H) and $0.64 \pm 0.01\,$GHz (V) for the X transition, and  $0.89 \pm 0.02\,$GHz (H) and $0.78 \pm 0.02\,$GHz (V) for the XX transition with an FSS of $2.28\ \pm 0.01\, \text{GHz}$.}
    \label{fig:figure2}
\end{figure}

To verify that the transfer process preserves the optical quality of the QD emission, we collect the photoluminescence spectrum of the NWQDs under pulsed above-band excitation before and after transfer (see Methods). One such spectrum for NWQD5 is shown in Figure \ref{fig:figure2}a, where we observe no change in the peak emission wavelength for three charge complexes as labeled in the top panel. We observe similar spectra before and after the transfer for three other NWQDs. Specifically, we observe no change in the emission wavelength for the neutral exciton (limited to the measurement precision of 15 GHz). Additionally, the spectral linewidths measured remain resolution-limited to the optical spectrometer. The data for all measured NWQDs is shown in SI Table \ref{table:all_transferred_nwqds}. Further, a scanning Fabry-Perot interferometer is used to obtain a higher-resolution measurement of the emission linewidth of the transferred NWQD5 (see Methods).

The raw data (black points) of this resolved photoluminescence are shown in Figure \ref{fig:figure2}b for the XX (top panel) and X (bottom panel) emission lines. 
The high-resolution spectrum reveals two peaks corresponding to the H and V dipoles of the XX/X emission arising from the FSS of the QD \cite{bayer2002fine}. The individual peaks are fit to a Lorentzian function shown as green (H-dipole) and orange (V-dipole) lines. The fitted linewidths are $0.71 \pm 0.01\,$GHz (H) and $0.64 \pm 0.01\,$GHz (V) for the X transition, and  $0.89 \pm 0.02\,$GHz (H) and $0.78 \pm 0.02\,$GHz (V) for the XX transition. Additionally, from the splitting between the H and V polarised peaks we extract an FSS of $2.28\ \pm 0.01\, \text{GHz}$. The sub-GHz Lorentzian linewidths are a good indication that the optical coherence of the photons emitted by the QD is maintained during the transfer as they are in a similar range to the previously reported values for NWQDs on growth-substrate \cite{Laferriere2023-kb}. Together, these spectral measurements indicate that the transfer process preserves the quality of the QD emission. 

\begin{figure}[h!p]
    \centering
    \begin{adjustbox}{center,width=0.5\linewidth}
        \includegraphics{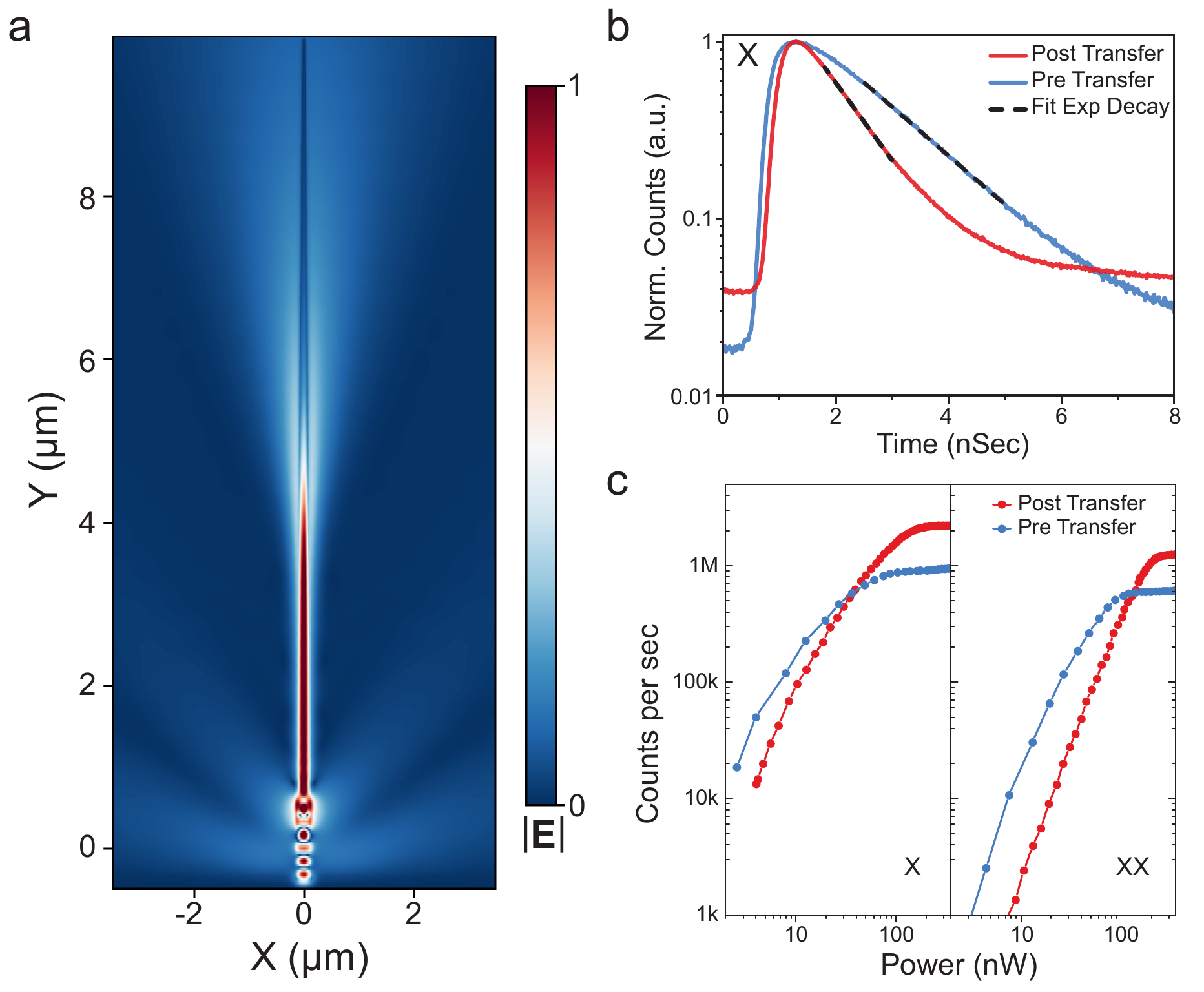}
    \end{adjustbox}
    \caption{NWQD on a Au/SiO$_2$ mirror. (a) 2D cross-section of the electric field intensity for a tapered nanowire on an Au/SiO$_2$ substrate from a 3D FDTD simulation. A weak cavity is formed between the bottom mirror substrate and the dipole as is evident by the standing wave in the intensity profile. The upward emission from the QD couples to a guided waveguide mode, visible as a continuous dark red region above the dipole, confined within the nanowire. As the mode propagates along the adiabatic taper, it gradually expands beyond the nanowire confinement, transitioning into a free-space mode with minimal back-reflections at a height of approximately $4$--$5\, \mu$m above the substrate.(b) Lifetime histograms of the neutral exciton before (blue) and after (red) the transfer of NWQD5 onto a Au/SiO$_2$ mirror (measured under above-band excitation). An exponential decay function is fit to the section shown as the dashed black line, which yields a lifetime of $1589 \pm 3 \,$ps before transfer and $1009 \pm 2 \,$ps after transfer. (c) Measured photon count rates of the exciton and biexciton transitions before (blue) and after (red) the NWQD transfer versus (above-band) excitation power. Overall, a $>2\times$ increase in brightness is observed for both the emission lines.}
    \label{fig:figure3}
\end{figure}

 To understand the electrodynamics of the NWQD on an Au/SiO$_2$ mirror substrate, we perform a Finite Difference Time Domain (FDTD) simulation of the system with realistic parameters. The dimensions of the nanowire are obtained using an SEM image and a classical dipole is placed around 1.5\,$\mu$m above the substrate to model the QD emission. The electric field intensity for an emission wavelength of 880\,nm is shown in Figure \ref{fig:figure3}a. The bottom mirror substrate reflects the fundamental waveguide mode back towards the QD with a modal reflectivity $>90\%$ \cite{Claudon2010-dp}. The modal back reflection results in the formation of a weak cavity in the waveguide as observed by intensity oscillations between the dipole and the bottom substrate indicating the presence of a standing wave. The upward emission couples efficiently to the propagating HE$_{11}$ mode and is observed as a continuous dark red region above the QD. After 3\,$\mu$m of propagation in the waveguide section, the adiabatic taper couples it into a free space Gaussian mode with negligible back reflections. The position of the QD along the height of the nanowire with respect to the bottom mirror determines the emission dynamics of the QD in this weak cavity. The expected Purcell enhancement of the radative lifetime $\gamma$ before and after transfer is given by $F_P = \gamma_{post}/\gamma_{pre} =\big(1+|r_m|\cos(\Phi)\big)$ where $|r_m|$ is the modal reflectivity of the bottom substrate and $\Phi$ represents the effective phase difference between light originating from the dipole and the path length of the emission reflected from the Au/SiO$_2$ mirror \cite{Friedler2009-uv}. A QD placed at the anti-node of the standing wave ($\Phi = 2\pi m$) results in the highest achievable Purcell factor of $F_P = 1.95$, for a realistic modal reflectivity of $r_m = 0.95$.

We experimentally determine the Purcell enhancement of all transferred nanowires by measuring the exciton radiative lifetime before and after transfer, which are given in SI Table \ref{table:all_transferred_nwqds}. Due to the variation in the exact location of the QD along the nanowire height ($\pm 100$nm) we observe both an increase and a decrease in the measured exciton radiative lifetime for different nanowires. For NWQD5 we observe a reduction in the exciton radiative lifetime by a factor of $1.58$ ($1589 \pm 3 \,$ps to $1009 \pm 2 \,$ps) as shown in Figure \ref{fig:figure3}b, indicating that the location of the QD is close to the anti-node of the weak cavity. Note that the measured linewidth is $\sim 4\times$ the Fourier limited linewidth of 0.16 GHz, thus confirming the high quality of the transferred NWQD. We also observe a similar reduction in the bi-exciton radiative lifetime as shown in SI Figure \ref{si_fig:xx_lifetime_au_mirror}. For the other 3 nanowires we conclude that the QD is closer to the node of the weak cavity, thus exhibiting an increase in the exciton radiative lifetime. Since the reflection from the Au/SiO$_2$ mirror substrate can mix the available guided modes of the nanowire waveguide \cite{jacobsen2023performance}, the weak cavity can couple the QD to higher order modes causing a non-Gaussian far-field emission. To confirm that the NWQD5 has a Gaussian far-field profile, we obtain k-space images of the QD emission of the NWQD before and after transfer as shown in SI Figure \ref{si_fig:D1_53_mode_image_Au_mirror_transfer}.  

Next, we systematically study the brightness of NWQD5. The measured count rates before and after transfer for the exciton and biexciton transitions are plotted in Figure \ref{fig:figure3}c. At the saturation power, the measured count rates of the exciton (biexciton) transition before and after transfer are 888 (590) kCounts/s and 2.12 (1.22) MCounts/s, respectively. As expected from the weak NWQD cavity, we observe an overall increase in the measured count rates.  To estimate the source collection efficiency, we calibrate the total end-to-end efficiency of the experimental setup (including the single-photon detector) to $\sim5\%$ (see \ref{si_sec:opt_eff}). The count rate of the exciton line from the NWQD5 on the gold mirror corresponds to a measured single-photon collection efficiency at the first lens of 53\% under above-band excitation. Note that these values exclude the phonon side band (PSB), which was filtered out by the measurement system (see Methods). To get an accurate assessment of the collection efficiency using above-band excitation, the count rates of both the neutral and charged exciton lines along with their respective PSB should be included \cite{Laferriere2022-aa}. This leads to a corrected collection efficiency at the first lens after transfer of 73\%, surpassing the 50\% limit for a NWQD on the growth substrate (see \ref{si_sec:opt_eff}).

We model the collection efficiency of the NWQD on a Au/SiO$_2$ mirror using a simple Fabry-Pérot model. In the model, the measured $F_P$ can be used to calculate the expected collection efficiency as given by \ref{eq:nwqd_purcell_collc_eff}. For NWQD5, the measured $F_P= 1.58$ yields an expected collection efficiency of 79\% as detailed in \ref{si_sec:opt_eff}, which is close to the measured value of 73\%. The small discrepancy between the measured and the calculated value can be attributed to factors such as error in the setup efficiency estimation and state preparation efficiency.     

\begin{figure}[h!]
    \centering
    \begin{adjustbox}{center,width=0.8\linewidth}
        \includegraphics{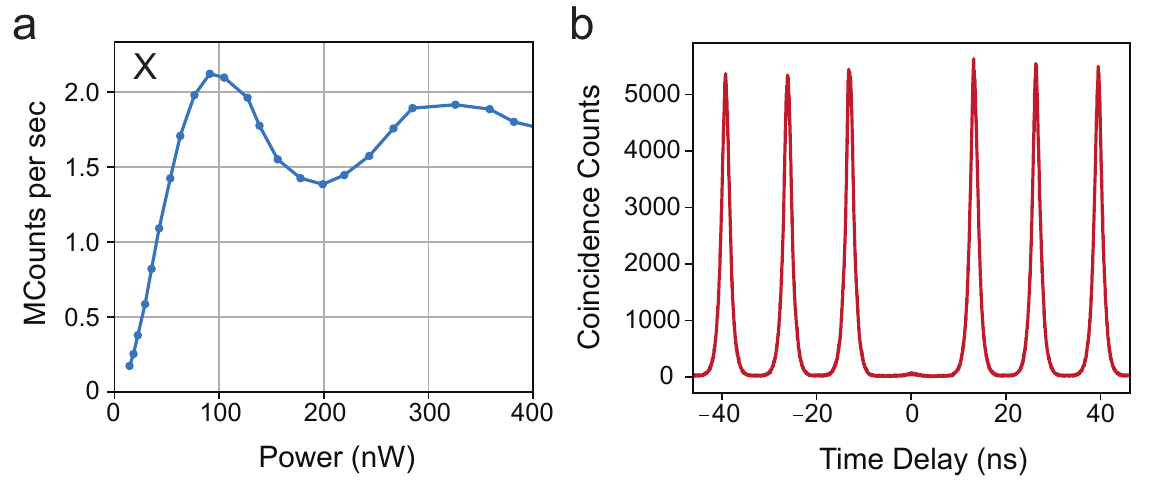}
    \end{adjustbox}    
    \caption{Two-photon excitation (TPE) of NWQD5 on a Au/SiO$_2$ mirror. (a) Measured photon count rates of the exciton versus excitation power showing clear Rabi oscillations. (b) Hanbury-Brown-Twiss (HBT) correlation histogram of the exciton at $\pi$-pulse ($\sim 100\,$nW) with a $g^{(2)}(0) = 0.0212\pm0.0001$.}
    \label{fig:figure4}
\end{figure}

To verify the measured count rates under above-band excitation and demonstrate low-multiphoton emission we conduct two-photon excitation (TPE) of NWQD5. Figure \ref{fig:figure4}a presents the detected count rate of exciton emission as a function of the excitation power under pulsed TPE. Due to the positive biexciton binding energy of the QD, we always get a significant amount of phonon-assisted resonant excitation of the exciton in addition to TPE, which lowers the visibility of the Rabi oscillations. The peak detected count rate at the $\pi$-pulse power is 2.12 MCounts/s, which is the same count rate obtained under above band excitation at saturation. At higher pulse area the phonon-assisted excitation dominates. For measured count rates of both X and XX emission with a higher pump power please refer to SI Figure \ref{si_fig:au_mirror_XX_X_rabi}. In Figure \ref{fig:figure4}b we show a standard Hanbury-Brown-Twiss (HBT) autocorrelation histogram of the exciton at $\pi$-pulse. We extract a $g^{(2)}(0) = 0.0212\pm0.0001$ indicating that NWQD5 after transfer still produces single photons with a low multi-photon contribution and that the excitation laser has been effectively extinguished (see Methods). Additionally, the long time delay HBT histogram (see SI Figure \ref{si_fig:blinking}) is used to extract the blinking ratio of $0.912\pm 0.002$ for NWQD5 under resonant excitation. The blinking ratio can be used to correct for the measured count rate, thus yielding a slightly higher collection efficiency of 75\% compared to above-band excitation. The implementation of TPE on NWQD5 demonstrates the feasibility of employing resonant excitation techniques with the NWQD on Au/SiO$_2$ substrate. For a NWQD with a negative biexciton binding energy, it will be possible to deterministically populate the biexciton under TPE and efficiently collect the entangled photons for a truly on-demand entangled photon source.

\bmhead{NWQD between Quadrupolar Gates}\hfill

\begin{figure}[h!]
    \centering
    \begin{adjustbox}{center,width=0.85\linewidth}
        \includegraphics{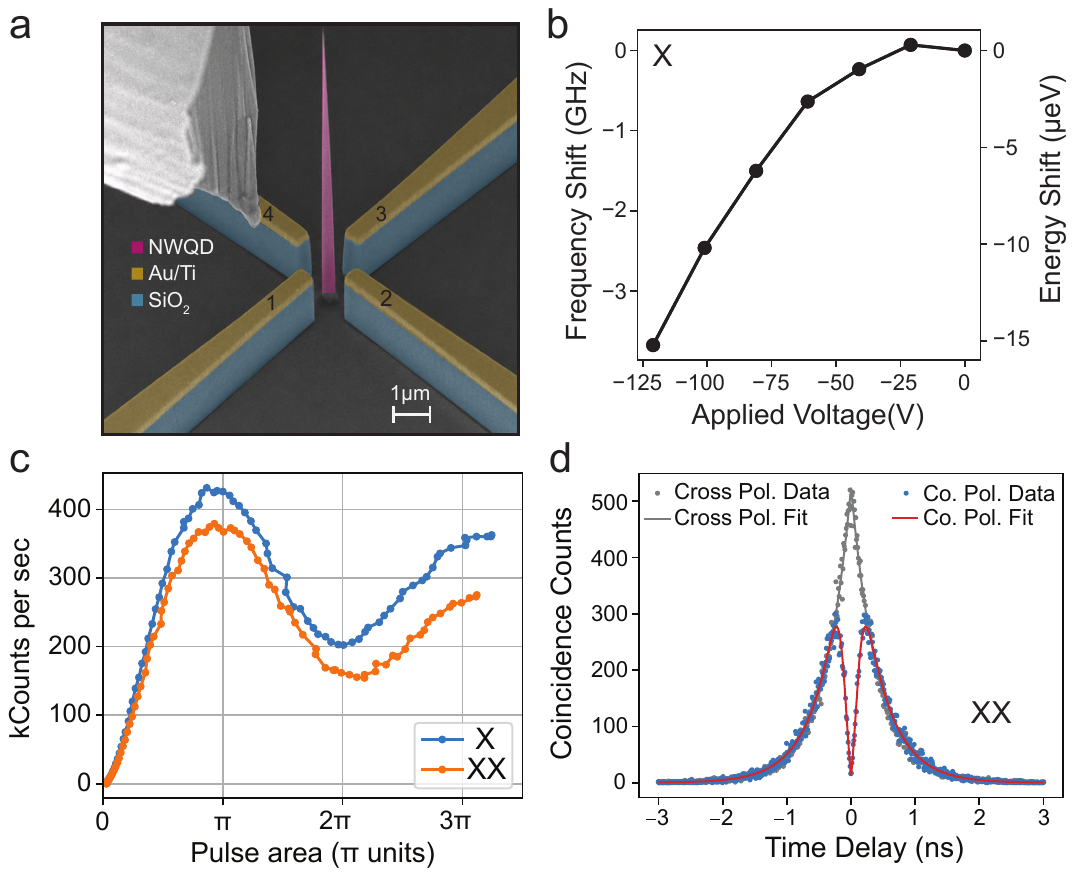}
    \end{adjustbox}    
    \caption{ (a) SEM image of the device with a NWQD (pink) transferred at the center of quadrupolar gates. The Ti/Au electrodes (yellow) are raised from the undoped Si substrate using a 1.5\,$\mu$m thick SiO$_2$ spacer (blue) to be in-plane with the QD in the nanowire. For the device measured, the gap between opposite electrodes at the center was 1\,$\mu$m. (b) Observed energy shift of the exciton emission as a function of an applied dipole electric field under cross-polarized resonance fluorescence. A maximum tuning range of 3.6\,GHz is obtained with an applied differential bias of 120\,V (raw spectra in SI Figure \ref{si_fig:tuning_raw_data}). (c) Measured Rabi oscillations of the exciton (blue) and biexciton (orange) entangled photon pair generated under two-photon excitation. (d) Cross- and co-polarized two-photon interference histograms of subsequently emitted biexciton photons generated at $\pi$-pulse (as shown in sub-figure c) with a raw TPI visibility of $24.5 \pm 0.5\, \%$.}
    \label{fig:figure5}
\end{figure}

For the second demonstration of the NWQD transfer, we use a target template with a quadrupole electrode geometry. We use this device geometry to apply a lateral electric field across the QD and tune its emission wavelength. An SEM of the device is shown in Figure \ref{fig:figure5}a. The quadrupole electrode geometry is designed as a symmetric cross with a 1.5\,$\mu m$ separation. The nanowire (pink) is placed at the centre of the cross using the NWQD transfer technique. The Ti/Au electrodes (yellow) sit $\sim 1.5\,\mu m$ above the Si substrate on an SiO$_2$ spacer (blue). The height of the electrodes is designed such that they will be approximately in-plane with the QD in the nanowire. See Methods for details on the nano-fabrication process of the quadrupole electrode target template. Similar to the first demonstration, we compare the photoluminescence spectrum under above-band illumination of the NWQD before and after transfer in SI Figure \ref{si_fig:tpe_quad_gates_spectra_pre_post}. We observe a small shift ($\sim0.2\,$nm) in the peak emission wavelengths for three prominent emission lines with no change in the linewidth (up to instrument resolution). This shift can be attributed to the altered electrostatic environment around the QD due to the presence of the electrostatic gates. 

With the quadrupolar electrode geometry, it is possible to apply a lateral dipole electric field to the QD. The dipole field shifts the emission wavelengths of the exciton complexes via the quantum confined Stark effect \cite{Reimer2008-wo}. To study the effect of an applied field on the QD we measure the neutral exciton resonance using cross-polarized resonance fluorescence. We scan a weak continuous wave (CW) laser with a narrow linewidth (10\,kHz) across the exciton resonance allowing us to accurately probe the center and the linewidth of the exciton emission (see Methods for the setup details). A dipolar electric field is applied to the QD where two adjacent gates (1 and 2 as labeled in Figure \ref{fig:figure5}a) are grounded and a voltage bias $V$ is applied to the opposite gates (3 and 4). The resulting lateral electric field distribution across the QD in the steady state is modeled in SI Figure \ref{si_fig:quad_gates_simulations}a. In Figure \ref{fig:figure5}b, we plot the measured shift in the emission wavelength as a function of the applied bias $V$ (see SI Figure \ref{si_fig:tuning_raw_data} for the raw spectra). We observe a 3.6\,GHz redshift of the emission peak for a maximum applied bias of -120\,V. The ability to tune the emission wavelength using an applied electric field is a significant step towards matching the emission wavelength of two separate quantum dots with applications in quantum networks, especially entanglement swapping between remote quantum dots.     

To demonstrate the viability of the quadrupole-gated NWQD device for generating entangled photon pairs we implement TPE of the biexciton state. The measured count rate of the biexciton (XX) and exciton (X) emission coupled into a single-mode fiber versus the pulse area is plotted in Figure \ref{fig:figure5}c with an applied $V$=\,0. We observe clear Rabi oscillations versus pulse power for both the biexciton and exciton emission, demonstrating the coherent nature of the TPE process. We then characterize the NWQD emission at $\pi$-pulse excitation. First, we measure the radiative lifetime of the XX and X emission as shown in SI Figure \ref{si_fig:tpe_quad_gates_lifetimes} and obtain $\tau_\text{XX} = 408.4\pm 0.1\,$ps and $\tau_\text{X} = 687\pm 4\,$ps. Second, the XX emission of the NWQD shows a very low multi-photon emission of $g^{(2)}(0) = 0.0021 \pm 0.0001$ and a blinking on/off ratio of $0.72$ as shown in SI Figure \ref{si_fig:quad_gates_hbt_blinking_XX}. 

Next, we characterize the coherence of two subsequently emitted photons at $\pi$-pulse excitation with a repetition rate of $76\,$ MHz using two-photon interferometry (TPI), see Methods for experimental setup details. The measured TPI histogram for the cross-polarized and co-polarized interfering photons (grey and blue dots) is displayed in Figure \ref{fig:figure5}d for an applied $V$=\,0. We find that the application of the dipole field does not induce any significant changes in the TPI histogram. The TPI visibility as a function of time window starting from the zero-time delay is plotted in SI Figure \ref{si_fig:windowed_hom}. The initial TPI visibility for the first $\pm 50\,$ps is greater than $90\%$ after which it considerably drops to $60\%$ by $\pm 200\,$ps emission window. The TPI visibility considering coincidences of the total emission (up to $\pm$5\,ns) is measured to be $24.5 \pm 0.5\, \%$. The central dip and adjacent peaks observed in the co-polarized histogram indicate optical decoherence arising from noise processes that reduce the indistinguishability of the emitted photons. The co(cross)-polarized histogram was fit (red(grey) lines) to the model developed by Kambs and Becher \cite{Kambs2018-di} to extract estimates for the magnitude of these dephasing processes and is shown in Figure \ref{fig:figure5}d. The model fits for two noise processes, namely, spectral diffusion and pure dephasing (see \ref{si_sec:tpi}). The fit indicates that the dominant noise mechanism is spectral diffusion corresponding to an inhomogeneous linewidth of $2.08\pm 0.04\,$GHz with a photon coherence time of \(228 \pm 6\)ps. The demonstrated wavelength tunability of 3.6 GHz in conjunction with the peak measured TPI visibility of $90\%$ establishes the feasibility of the interfacing such a device with other modules of a quantum network. Possibilities include interfacing with another NWQD entangled photon source, atomic-clouds for quantum memories and ions/netural atoms qubits of a quantum computer. 

To improve the device performance beyond the 3.6 GHz tuning range and achieve higher photon indistinguishability for longer photon time delays, novel device geometries must be explored. It has been previously shown that for field strengths of tens of kV/cm of an applied lateral field to a QD, shifts on the order of hundreds of GHz have been observed \cite{Kaniber2011-kt, Reimer2008-wo}. For simulations shown in SI Figure \ref{si_fig:quad_gates_simulations}b, our device should show a similar magnitude of tuning for an applied field that is $100\times$ lower than that applied to the device in Figure \ref{fig:figure5}b. This suggests that the QD is being screened from the applied electric field, which could be due to two plausible causes: dopants in the InP nanowire or surface-states at the nanowire-air interface. Indeed, if we include dopants in the nanowire for the simulations, the model shows reduced tunability as shown in SI Figure \ref{si_fig:quad_gates_simulations}b. The reduced photon indistinguishability primarily caused due to spectral wandering can be eliminated by stabilizing the charge environment around the QD. Such an approach has been shown to achieve $>95\%$ TPI visibility for long time delays and Fourier limited linewidths \cite{tomm2021bright}. To achieve such long time scale QD coherence, the template and the NWQD must be engineered such that the QD is located close to a Fermi-sea with an appropriate tunneling barrier. Establishing such a tunneling barrier will also allow for charge state control of the QD.      

\section*{Summary}

In this work, we have presented a deterministic pick-and-place transfer technique utilizing nano-gluing to move NWQDs grown in site-selected manner onto arbitrary pre-fabricated templates. Critically, the transfer procedure maintains the vertical geometry of the NWQDs, allowing for high efficiency coupling of the nanowire waveguide into free space. We have shown that the transfer process maintains the high optical quality of the QDs, including their measured linewidth, low-multiphoton emission probability, and photon indistinguishability. 

Using this transfer technique we have implemented two device geometries to enhance the photon extraction efficiency and enable electrical tuning of the emission wavelength. In the first geometry, we implemented a bottom Au/SiO$_2$ mirror and demonstrated a weak nanowire cavity with single-photon extraction efficiency of 75\%, accompanied by a Purcell enhancement of 1.58. In the second geometry, quadrupole electrodes were implemented around the QD and were used to controllably induce a shift in the emission wavelength over a range of 3.6 GHz. 

Future device designs could combine the high extraction efficiency with emission-wavelength tunability and charge state control in a single NWQD device. These devices will be aimed to improve the coherence of the optical emission \cite{Zhai2020-yh, Kuhlmann2015-dd} and also tune the exciton FSS \cite{Zeeshan2019-na} to realize an ideal on-demand entangled photon pair source. The versatility of the EBID assisted transfer process can also be exploited beyond the vertical-to-vertical transfer regime. For example, NWQDs can be repeatably transferred in a `vertical'-to-`lying-down' orientation onto pre-fabricated photonic integrated circuits \cite{Mnaymneh2020-lv}. We further envision that the process developed in this work will be used to integrate NWQDs with microwave circuitry for advanced spin control and with optomechanical resonators for quantum transduction. Such capabilities are important building blocks for the generation of high-fidelity photonic cluster states and the realization of quantum memories based on semiconductor quantum dots, paving the way towards next-generation quantum technologies.

\bmhead{Acknowledgments}
We acknowledge Travis Casagrande for providing SEM images used in Figure 1. We acknowledge François Sfigakis for useful discussions on nanofabrication of the quadrupole template and Catalina Gutiérrez for their help in building the TPI setup.
\backmatter
\nolinenumbers
\clearpage
\section*{Methods}

\setlength{\parindent}{0pt}

\textbf{Fabrication of the Au/SiO$_{2}$ Mirror}

A 10/100 nm layer of Ti/Au was deposited by electron-beam evaporation on an undoped Si wafer.
After the Au deposition, 10 nm of SiO$_{2}$ was deposited with plasma-assisted atomic layer deposition with bis(tertiary-butylamino)silane and O$_{2}$ precursors at 300 $^{\circ}$C. 
The free-space reflectivity of the mirror substrate was measured to be 97.9\%.
The purpose of the SiO$_{2}$ coating is to increase the modal reflectivity of the HE$_{11}$ mode \cite{Claudon2013-wn, Friedler2009-uv}.
\vspace{1\baselineskip}

\textbf{Fabrication of the Quadrupole Gates}

1.4 $\mu$m of SiO$_{2}$ was deposited on an undoped Si wafer, using plasma-enhanced chemical vapor deposition with SiH$_{4}$ and N$_{2}$O precursors at 330 $^{\circ}$C. 
The quadrupole gates were defined by electron-beam lithography and a Ti/Au/Cr (20/200/180 nm) layer stack was deposited by electron-beam evaporation. 
Interconnects and bond pads were then patterned with optical lithography and a Ti/Au/Cr (20/100/180 nm) layer stack was deposited by electron beam evaporation. 
The sample was then etched using reactive ion etching with C$_{4}$F$_{8}$ and O$_{2}$ to anisotropically etch the SiO$_{2}$ around the gates, leaving the gates elevated with respect to the exposed substrate. 
The Cr layer acts as an etch mask with good selectivity with respect to SiO$_{2}$. 
The remaining Cr after the SiO$_{2}$ etch was removed with Chromium Etchant 1020 to allow for wirebonding to the Au pads.
\vspace{1\baselineskip}

\textbf{NWQD Excitation and Collection}

The NWQD on a Au/SiO$_{2}$ substrate was cooled in an attoDRY 800 cryostat to 4.5\,K and the NWQD between quadrupole gates was cooled in an attoDRY 2100 cryostat to 1.7\,K. In both cases, a cryogenic objective was used to focus the input excitation light into the nanowire and and collect the QD emission. A 70:30 (T:R) beam splitter was used to separate the input-excitation (R) and output-collection (T) paths. The above-band excitation was conducted using a pulsed Ti:Sapphire laser at $830\,$nm wavelength and $3\,$ps pulse width. For above-band excitation, the saturation power was defined by the input power required to obtain a count rate plateau of the exciton emission. For two-photon excitation (TPE), the pulsed Ti:Sapphire laser with $3\,$ps pulse width was set to a wavelength halfway between the biexciton and exciton transitions. The pulses were then elongated in time to $\sim 20\,$ps using a 4F pulse shaper. The pulse shaper consists of two $1200\,$g/mm gratings and a lenses with focal length of $150\,$mm.
Two $0.1\,$nm bandwidth volume Bragg grating notch filters were used to reject the excitation laser in the collection path.
\vspace{1\baselineskip}

\textbf{NWQD on Gold Characterization}

To record the optical spectrum in Figure \ref{fig:figure1}a, the collected photons from the NWQD were directed via free-space to a $1200\,$g/mm spectrometer with a resolution of $10-20\,$GHz (Princeton Instruments HRS 750 with PIXIS CMOS camera).  
To record the data in Figure \ref{fig:figure1}(b,c), a free-space single photon avalanche diode (SPAD) placed at the output of the exit slit of the spectrometer was used to detect the single photons of the zero-phonon-line for each optical transition.
See Supplementary Section \ref{si_sec:opt_eff} for the calculations of the total path efficiency.
The radiative lifetime histogram was produced by correlating the zero-phonon-line emission for a given transition with the pulse train of the excitation laser. 
The TPE Rabi oscillations were recorded using the same setup.
To measure the HBT autocorrelation histogram, the NWQD emission was coupled to single-mode fibre (780HP) and sent to a 50:50 fibre-based beam splitter, whose outputs were detected on two fibre-coupled SPADs.
The high resolution linewidth scans in Figure \ref{fig:figure1}b were done using a Thorlabs (SA210-8B) scanning Fabry-Perot interferometer (FPI).
The nominal free-spectral range (FSR) and finesse of the FPI are 10GHz and 180, respectively.
A Qontrol Ltd. Q8iv programmable voltage source was used to drive the piezo actuators inside the FPI to scan a full FSR with a step size of 30$\,$MHz.
\vspace{1\baselineskip}

\textbf{Resonant Drive Linewidth Scans}

A Toptica CTL 900 tunable diode laser with a nominal linewidth of $\sim 10\,$kHz was used to perform the linewidth scans of the neutral exciton emission line in Figure \ref{fig:figure5}b.
The intensity of the input drive laser was set to around a mean photon number per lifetime of 0.1 to ensure no power broadening was present.
A typical cross-polarized excitation-collection configuration was used reject background laser and obtain a clean resonance fluorescence (RF) signal.       
The RF signal from the NWQD was collected into SMF and sent to Single Quantum B.V. superconducting nanowire single photon detectors (SNSPDs) with a FWHM timing-jitter of $8\,$ps and an efficiency of $\sim 85\%$.
The lineshape is fitted to a Gaussian function to extract the central frequency and the linewidth of the Quantum emitter(FWHM).
Two Keithley 2450 source meters were used to apply bias to the quadrupole gates around the NWQD.

\vspace{1\baselineskip}

\textbf{Two-Photon Interference Measurement}

Given the pulse separation of $\sim 13\,$ns for the excitation laser, an asymmetric Mach-Zehnder Interferometer (MZI) was build such that the total delay between the paths was $\sim 13\,$ns. 
Thus, subsequently emitted photons overlap in time at a 50:50 beam splitter and undergo two-photon interference.
Single photon correlations are detected between the two output ports of the beam-splitter using SNSPDs and a time tagger (Swabian Instruments). 
The emission line sent to the MZI was selected using a fibre-coupled tunable bandpass filter with a FWHM of $\sim0.07\,$nm or $\sim30\,$GHz and it was polarised using a nano-particle polariser.
A quarter and half waveplate in one path of the MZI were used to record the co-polarised or cross-polarised histograms.
\vspace{1\baselineskip}

\section*{Declarations}

\bmhead{Funding} 
M.E.R. and D.D. would like to acknowledge financial support from NSERC Quantum Alliance, NSERC Discovery, and NRC Quantum Sensing Program, which funded part of this research.
The University of Waterloo's QNFCF facility was used for this work. This infrastructure would not be possible without the significant contributions of CFREF-TQT, CFI, ISED, the Ontario Ministry of Research \& Innovation and Mike \& Ophelia Lazaridis. Their support is gratefully acknowledged.

\bmhead{Conflict of interest/Competing interests}

\bmhead{Author contribution}

M.E.R. conceived and led the project.
S.V.G and S.G conceptualized and initiated the NWQD transfer process with input from G.H.
G.H., T.P., and M.P. developed the NWQD transfer process.  
M.P. and T.P. designed experiments with inputs from M.E.R., S.G., D.D, and P.P. 
M.J., D.D., and P.P. grew the NWQDs.
S.V-G. and S.R.H. fabricated the Au/SiO${_2}$ template. 
S.R.H. fabricated the quadrupole gate template with inputs from T.P. and D.D. 
G.H. performed the NWQD transfer onto the templates with input from T.P. 
M.P. and T.P. conducted the optical measurements for NWQD on the Au/SiO${_2}$ template.
M.P., S.G., and T.P. conducted the optical measurements for NWQD between quadrupole gates.
M.P., A.D., and T.P. performed the data analysis. 
A.D. conducted the electrostatic simulations with inputs from M.E.R., T.P., and M.P. 
M.P., T.P., and S.G. wrote the manuscript with input from all authors.

\noindent

\bigskip

\bibliography{paperpile, bibliography}

\clearpage
\setcounter{section}{0}
\setcounter{equation}{0}
\setcounter{figure}{0}
\pagenumbering{arabic}
\renewcommand{\thepage}{S\arabic{page}} 
\renewcommand{\thesection}{SI Section \arabic{section}}
\renewcommand{\thesubsection}{SI Section \arabic{section}.\arabic{subsection}}  
\renewcommand{\thetable}{\arabic{table}}  
\renewcommand{\thefigure}{\arabic{figure}}
\renewcommand{\theequation}{SI Eqn. \arabic{equation}}

\captionsetup[figure]{name=SI Figure}
\captionsetup[table]{name=SI Table}


{\centering{\Large \textbf{Supplementary Information}}}

\section{Details of nanowire detachment} \label{si_sec:nwqds_on_mirror}

\begin{figure}[h!]
    \centering
    \includegraphics[width=1\linewidth]{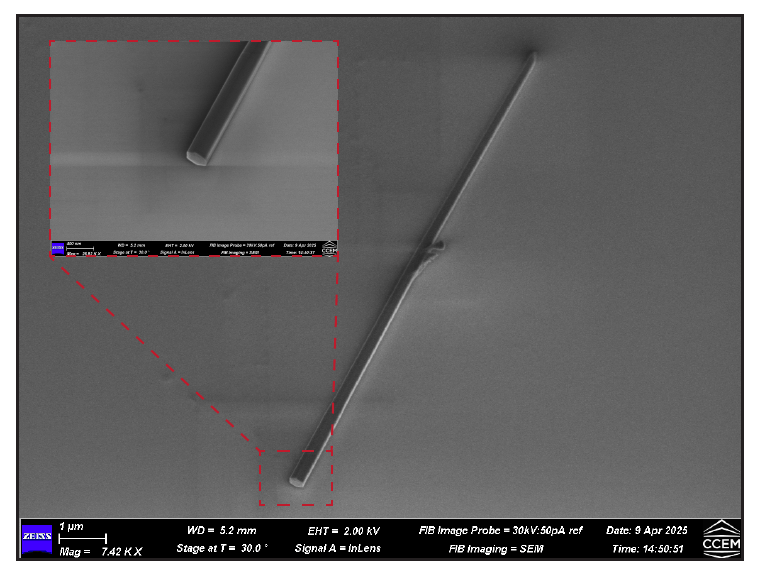}
    \caption{SEM image of a detached nanowire lying down on the substrate. A zoomed in inset shows a flat bottom facet with the hexagonal cross-section of the wurtzite nanowire. The wurtzite nanowire cleaves along it's lattice plane due to the diameter mismatch at it's base on the growth substrate. This mismatch results form overgrowing the 100 nm diameter oxide hole on the growth substrate.}
    \label{si_fig:nwqds_flat_cleave}
\end{figure}

\clearpage

\section{Details of NWQDs on Au/SiO${_2}$} \label{si_sec:nwqds_on_mirror}

\begin{figure}[h!]
    \centering
    \includegraphics[width=1\linewidth]{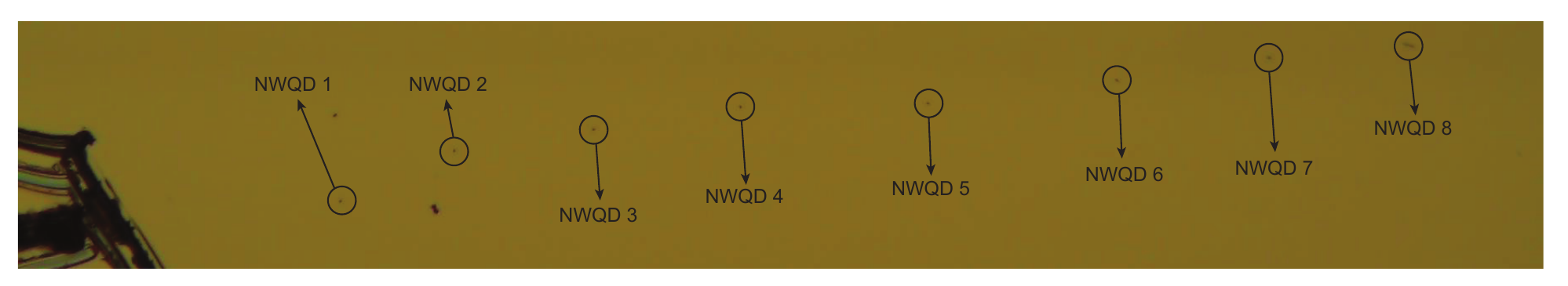}
    \caption{Optical microscope image (100x) of the eight NWQDs transferred onto the Au/SiO${_2}$ mirror template. The labeled NWQD5 is the device under study in the main text.}
    \label{si_fig:nwqds_on_mirror}
\end{figure}

In total, eight NWQDs were transferred from the growth substrate onto the Au/SiO${_2}$ mirror.
An optical microscope image of the final device is shown in SI Figure \ref{si_fig:nwqds_on_mirror}.
Of these eight, five (NWQD4-8) had been pre-characterized on the growth substrate using above-band excitation. The pre-characterization included, power dependent emission spectrum, power dependent count rates and lifetimes of the three major complexes (neutral-, bi-, and charged excitons).
The optical properties of the NWQD emission before and after transfer are compared in SI Table \ref{table:all_transferred_nwqds}.
Only NWQD5 showed a decrease in lifetime and an increase in the detected count rate for all three exciton lines. 
All other NWQDs showed a longer lifetime after transfer; however, there was no change in their optical linewidth on the spectrometer. 
This suggests that there is no degradation of the QD and that the QD is simply closer to the node of the electric field reflected by the mirror, causing the lifetime to increase. An additional three (NWQD1-3) were transferred but were not pre-characterized before the transfer and were therefore omitted from the analysis. NWQD8 was also omitted from the analysis since the transfer process was not completely successful and the transferred NW was not perfectly vertical.     

\begin{table}[h]
\centering
\normalsize
\begin{tabular}{@{}l l l l@{}}
\toprule
NWQD & $\lambda$ in nm& $\Delta\lambda_\text{FWHM}$ in pm & $\tau$ in ps\\
\midrule
NWQD4 & 880.307 / 880.326 & 28.2 (0.3) / 26.9 (0.5) & 965(1) / 1467(5) \\
NWQD5 & 878.279 / 878.296 & 25.8 (0.4) / 26.0 (0.8) & 1589(3) / 1009(2) \\
NWQD6 & 880.090 / 880.129 & 29 (3) / 29 (5) & 1112(4) / 2300(10) \\
NWQD7 & 884.843 / 884.84 & 28 (13) / 36 (15) & 977(3) / 1736(5)\\
\bottomrule
\end{tabular}
\caption{Table of the neutral exciton emission wavelength $(\lambda)$, emission linewidth (measured on a grating spectrometer) $(\Delta\lambda_\text{FWHM})$ and emission lifetime $(\tau)$, pre-transfer and post-transfer (pre/post) respectively. }
\label{table:all_transferred_nwqds}
\end{table}
\vspace{2cm}

\section{Efficiency Calculations} \label{si_sec:opt_eff}

\textbf{Experimental Setup Efficiency.}

To get an accurate estimate of the first lens efficiency of the QD emission, for NWQD on Au/SiO${_2}$ mirror, we first characterize the optical path efficiency and single photon detection efficiency.
This was done by measuring the transmission and detection efficiency for each of the constituent parts listed in SI Table \ref{table:optical_setup_eff} using a power meter and a pulsed laser tuned to the NWQD emission wavelength of $880\,$nm.

\begin{table}[h]
\centering
\normalsize
\begin{tabular}{@{}l r@{}}
\toprule
Component & Transmission/Detection efficiency \\
\midrule
Cryostat (objective, windows, mirror) & 0.77 \\
Beam splitter & 0.65 \\
Free-space path to spectrometer & 0.84 \\
Spectrometer & 0.28 \\
SPAD & 0.46 \\
\midrule
Total & 5.3\% \\
\bottomrule
\end{tabular}
\caption{Experimentally measured efficiency values for the various components used in the experimental setup to characterise the QD emission of the NWQD on  Au/SiO${_2}$.}
\label{table:optical_setup_eff}
\end{table}
\vspace{-1cm}

As discussed in the main text, to get an accurate estimate of the total NWQD collection efficiency, corrections for the phonon side-band and the non-linear detection response of the SPAD at high count rates (>1 MCounts/s) should be included. These values are listed in SI Table \ref{table:eff_params}.

\begin{table}[h]
\centering
\normalsize
\begin{tabular}{@{}l r l@{}}
\toprule
Detector Non-linearity at 1Mc/s & $c_\text{D} = 1.02$ \\
Detector Non-linearity at 2Mc/s & $c_\text{D} = 1.08$ \\
Phonon-side band fraction \cite{Laferriere2022-aa} & $c_\text{PSB} = 1.2$ \\
Total detection efficiency & $\eta_\text{opt} = 5.3\%$ \\
Laser Repetition Rate & $f = 76.2\,$MHz \\
\bottomrule
\end{tabular}
\caption{Experimental parameters and correction factors used to estimate the first lens collection efficiency of the NWQD emission.}
\label{table:eff_params}
\end{table}
\vspace{-1.5cm}

Finally, the corrected and uncorrected count rates are given by 
\begin{equation}
    \eta_\text{uncorr} = \frac{N}{f}\frac{1}{\eta_\text{opt}} \quad \text{and} \quad \eta_{corr} = \frac{N}{f}\frac{c_D\cdot c_\text{PSB}}{\eta_\text{opt}} \,
\end{equation}
respectively, where $c_D$ is the detector non-linear correction factor at 1\,MCounts/s or 2\,MCounts/s depending on the count rate $N$.
The values for the corrected and uncorrected efficiencies calculated using the two equations above and the values listed in SI Table \ref{table:eff_params} are given in SI Table \ref{table:nwqd_fl_eff}.

\begin{table}[h]
\centering
\normalsize
\begin{tabular}{@{}c c c c@{}}
\toprule
Device and Emission Lines & Measured Count Rate & $\eta_\text{uncorr}$ & $\eta_\text{corr}$ \\
\midrule
On Au $X^0$ + $X^-$ &	$2.27\,\times 10^{6}$ & $56\%$ & $73\%$ \\ 
On Au $X^0$ & $2.13\,\times 10^{6}$ & $53\%$ & $68\%$\\
On Substrate $X^0$ + $X^-$ & $9.61\,\times10^{5}$ & $24\%$ & $29\%$\\
On Substrate $X^0$ & $8.88\,\times10^{5}$ & $22\%$ & $27\%$\\
\bottomrule
\end{tabular}
\caption{Experimentally measured count rates used to estimate the first lens collection efficiency of the NWQD on a Au/SiO${_2}$ mirror and on the growth substrate.}
\label{table:nwqd_fl_eff}
\end{table}
\vspace{-1.5cm}

\textbf{Theoretical Extraction Efficiency of NWQD on Au/SiO${_2}$ substrate}

The overall photon extraction efficiency of the photonic nanowire on a Au/SiO${_2}$ substrate with an adiabatic taper on the other end, is estimated by \cite{Claudon2013-wn}
\begin{equation}
   \eta = \frac{T_{\alpha}}{2}\frac{\beta(1+|r_m|^2 + 2\cos(\Phi)|r_m|)}{1+\beta\cos(\Phi)|r_m|}
   \label{eq:nwqd_purcell_collc_eff}
\end{equation}
where $\beta$ is the coupling strength of the QD to an infinitely long nanowire waveguide, $r_m$ is the modal reflectivity of the bottom mirror, $T_{\alpha}$ is the transmission ratio through the top taper and $\Phi$ is the phase difference between the dipole and the reflected field from the mirror. An experimentally achievable peak efficiency of $\eta > 0.95$ is possible when, $\Phi = 2\pi m$, $|r_m|=0.95, \beta = 0.96$ and $T_{\alpha<1} = 1$. For our device, we can accurately estimate that $\beta \approx 0.9$ from the measured diameter of the nanowire using SEM imaging. The SEM image also allows for an accurate measurement of the top taper angle ($\sim 3^\degree$) which implies a $T_\alpha = 0.9$. The $|r_m|$ can be assumed to be 0.9, factoring fabrication imperfections. We can then find $\Phi$ from the measured Purcell enhancement using the expression for $F_P$ given in the main text. Using these parameters we estimate the extraction efficiency of $\eta\approx79\%$, which is close to the measured value of 75\%. 

\vspace{1\baselineskip}

\clearpage

\section{Additional Data for NWQD5}
\subsection{Biexciton Radiative Lifetime}
\begin{figure}[!h]
    \centering
    \includegraphics[width=0.65\linewidth]{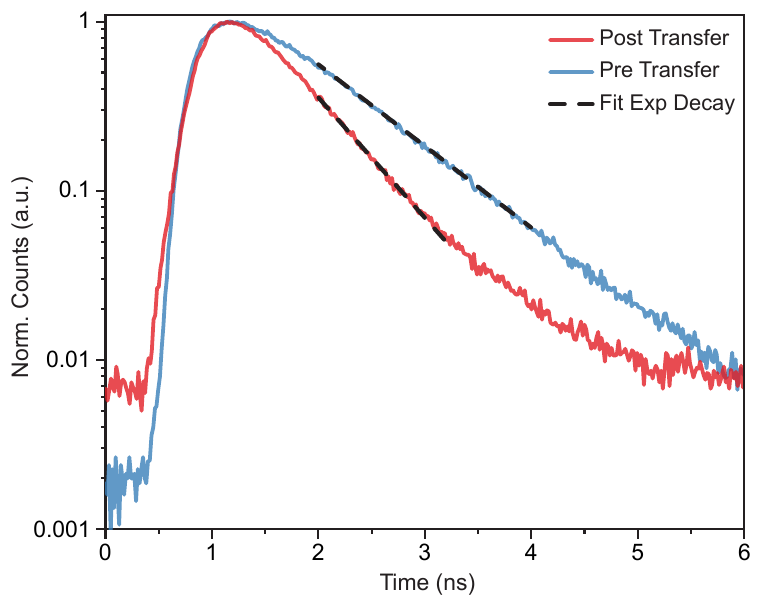}
    \caption{Lifetime histogram of the biexciton transition of NWQD5 using above-band excitation before (blue) and after (red) transfer onto the Au/SiO${_2}$ mirror. The lifetimes before and after transfer are $921\pm3\,$ps and $626\pm3\,$ps respectively, yielding a Purcell factor of $1.47$ . }
    \label{si_fig:xx_lifetime_au_mirror}
\end{figure}
\clearpage

\subsection{Blinking of the Exciton Transition}
\begin{figure}[!h]
    \centering
    \includegraphics[width=0.65\linewidth]{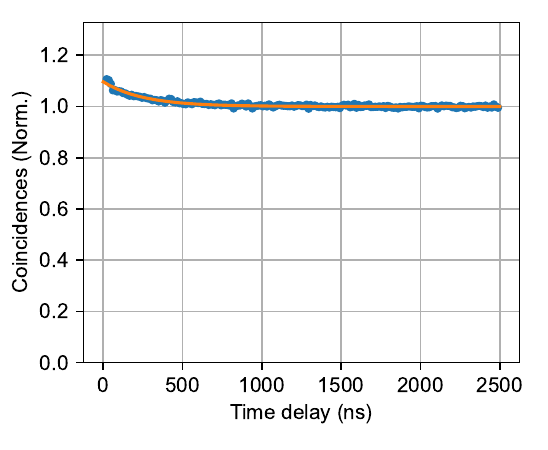}
    \caption{Heights of the HBT correlation histogram peaks from NWQD5 for time delays of up to 2.5$\,\mu$s (blue dots) and the fit blinking function (orange). The extracted blinking ratio and correlation-time is $0.912\pm0.002$ and $0.25\pm0.01\,\mu$s, respectively.}
    \label{si_fig:blinking}
\end{figure}

\clearpage
\subsection{Rabi Oscillations}
\begin{figure}[!h]
    \centering
    \includegraphics[width=0.65\linewidth]{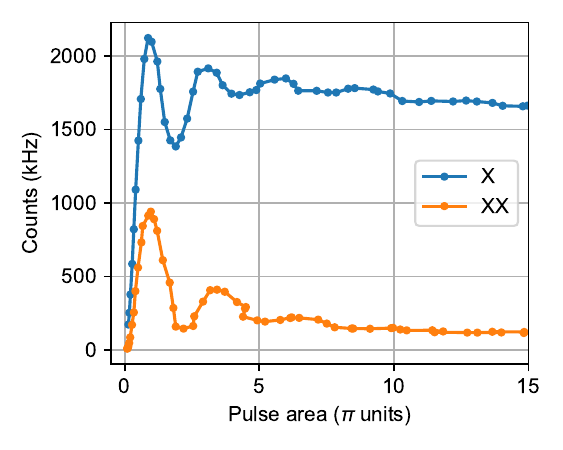}
    \caption{Measured count rates of the exciton (X) and biexciton (XX) transitions under two-photon excitation for NWQD5. As a result of large positive binding energy of the quantum dot, the two-photon resonance overlaps with the frequency band for photon-assisted excitation of the exciton. As a result, the Rabi oscillations are severely damped for both transitions and for large pulse intensities, phonon assisted excitation of the exciton dominates, heavily suppressing the generation of the biexciton. }
    \label{si_fig:au_mirror_XX_X_rabi}
\end{figure}

\clearpage

\subsection{K-Space Images of Emission Mode}
\begin{figure}[!h]
    \centering
    \includegraphics[width=1\linewidth]{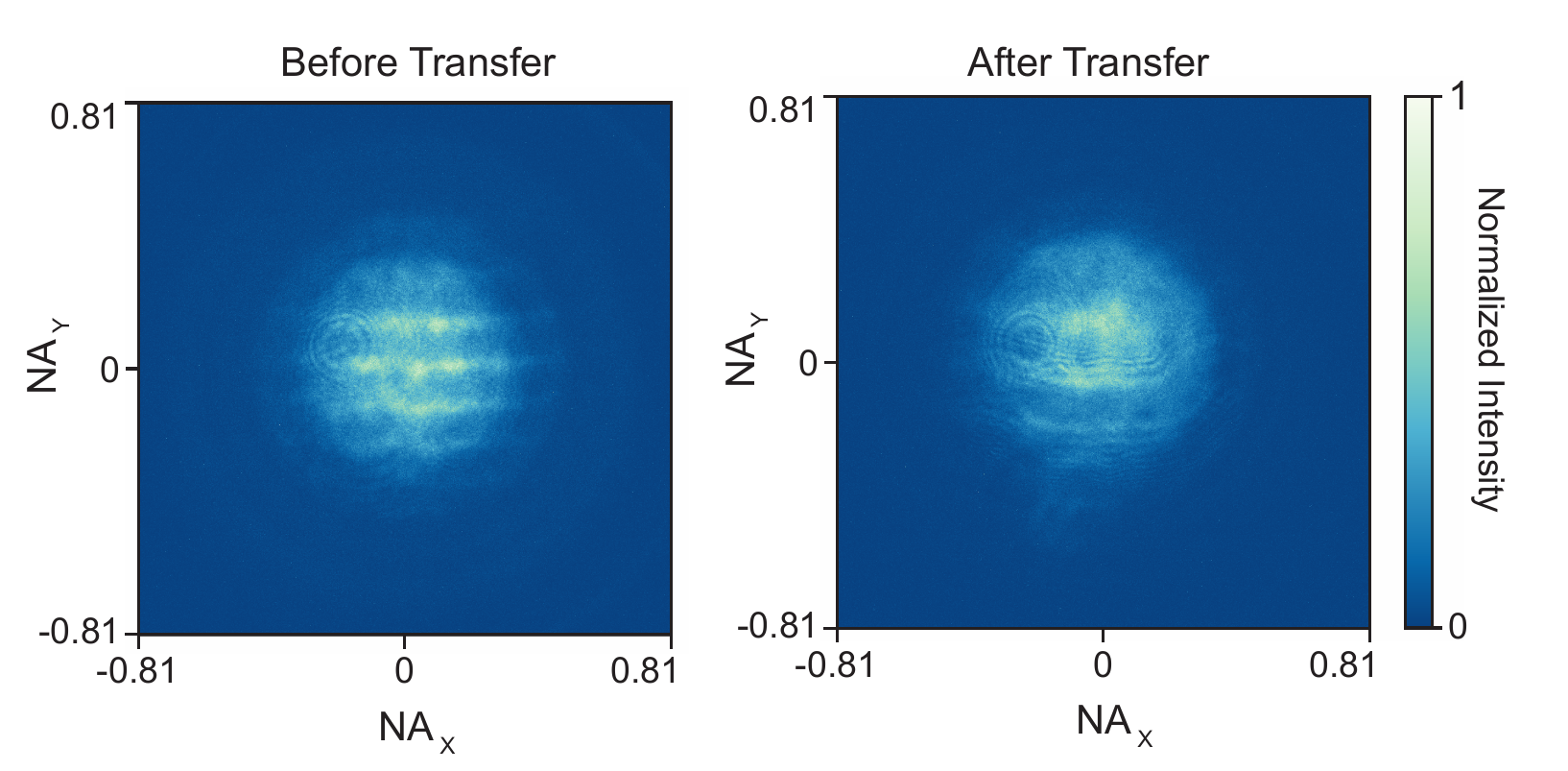}
    \caption{K-space distribution of the QD emission for NWQD5 before and after transfer.}
    \label{si_fig:D1_53_mode_image_Au_mirror_transfer}
\end{figure}

\clearpage

\begin{adjustwidth}{-1cm}{-1cm}
\section{Additional Data for NWQD between Quadrupole-Gates} \label{si_sec:additional_quad_data}
\end{adjustwidth}

\subsection{Above-band Emission Spectra}
\begin{figure}[!ht]
    \centering
    \includegraphics[width=0.65\linewidth]{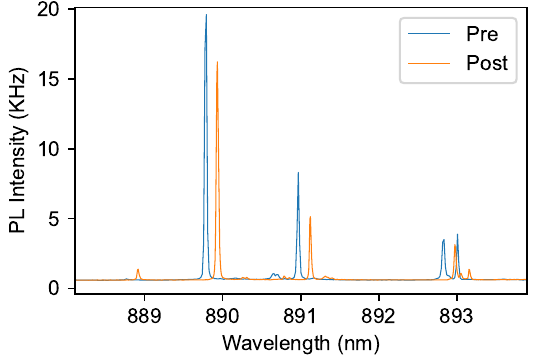}
    \caption{Emission spectra before (blue) and after (yellow) NWQD transfer with no applied bias using above-band excitation. The central wavelengths $(\lambda)$ and linewidths $(\Delta)$ before and after transfer are: $\lambda_\text{pre} = 889.785$ nm with $\Delta_\text{pre} = 28.4\pm0.4\,$pm and $\lambda_\text{post} = 889.933$ nm with $\Delta_\text{post} = 31\pm1\,$pm. }
    \label{si_fig:tpe_quad_gates_spectra_pre_post}
\end{figure}

\clearpage

\subsection{Wavelength Tuning of QD Emission}
\begin{figure}[!ht]
    \centering
    \includegraphics[width=1\linewidth]{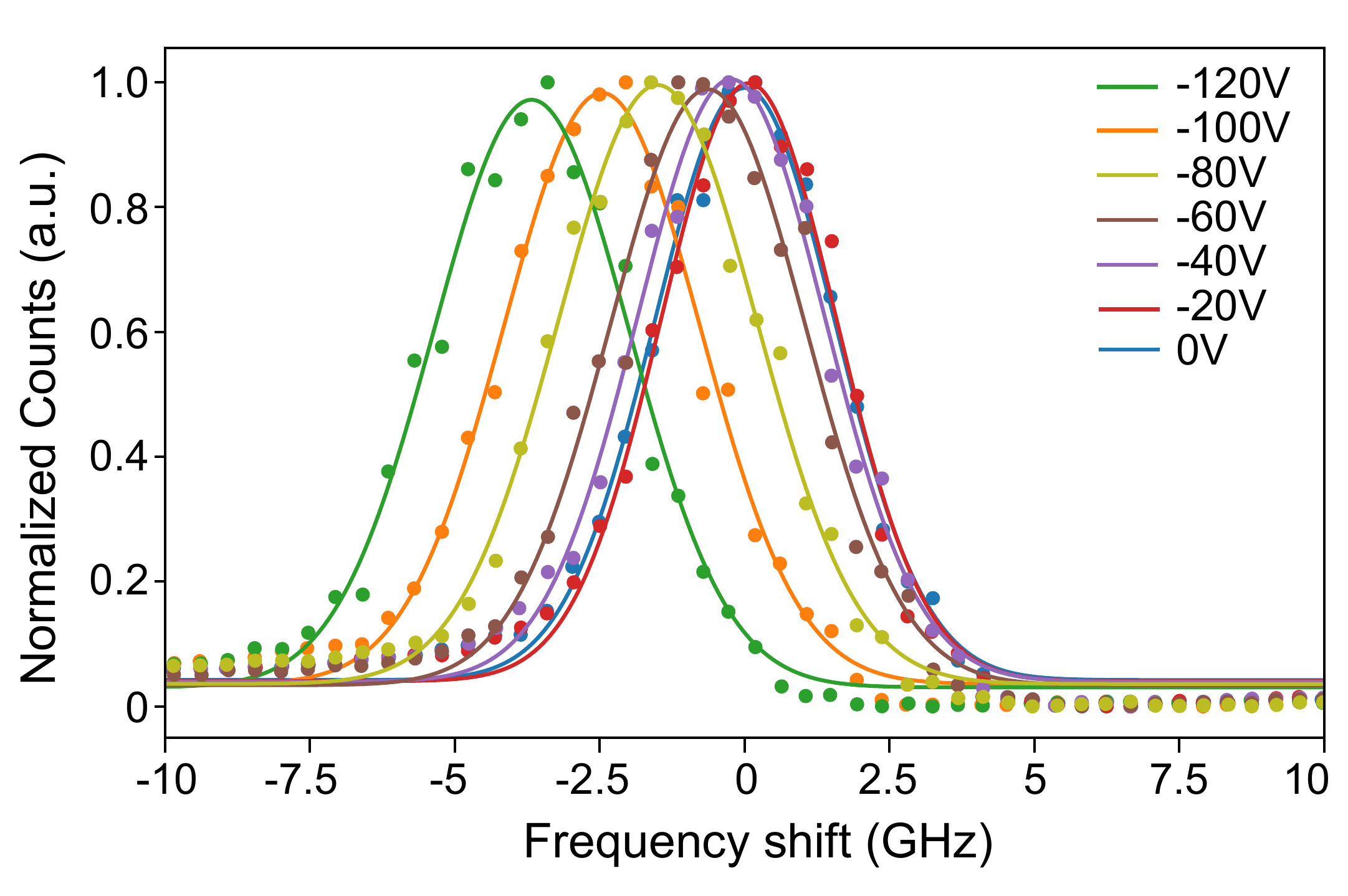}
    \caption{Measured resonant spectrum of the exciton emission under an applied dipole electric field. The spectrum is obtained under a cross-polarized excitation-collection scheme with a weak (less than 0.1 photons per lifetime) narrow linewidth (10\,kHz) excitation laser. A line-scan spectrum is obtained by sweeping the laser wavelength across the exciton transition for various applied biases ranging from 0V to 120V. The raw data is plotted as a function of detuning of the laser from the center of the exciton resonance at 0V as colored dots. The data is fit to a Gaussian function (solid lines) to determine the shift in the center frequency, which is shown in Figure \ref{fig:figure5}.}
    \label{si_fig:tuning_raw_data}
\end{figure}

\clearpage

\subsection{Radiative Lifetimes under TPE}
\begin{figure}[H]
    \centering
    \includegraphics[width=1\linewidth]{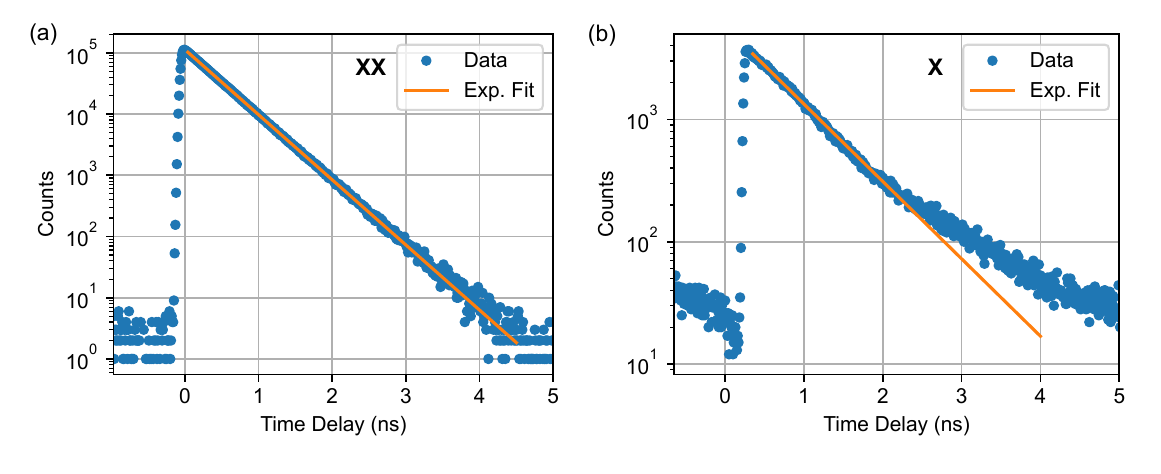}
    \caption{Radiative lifetime histograms of the NWQD (blue dots) with zero applied bias under TPE. From an exponential fit (yellow line)to the histogram we extract: $\tau_\text{XX} = 408.4\pm 0.1\,$ps and $\tau_\text{X} = 687\pm 4\,$ps}
    \label{si_fig:tpe_quad_gates_lifetimes}
\end{figure}

\subsection{HBT and Blinking}
\begin{figure}[H]
    \centering
    \includegraphics[width=1\linewidth]{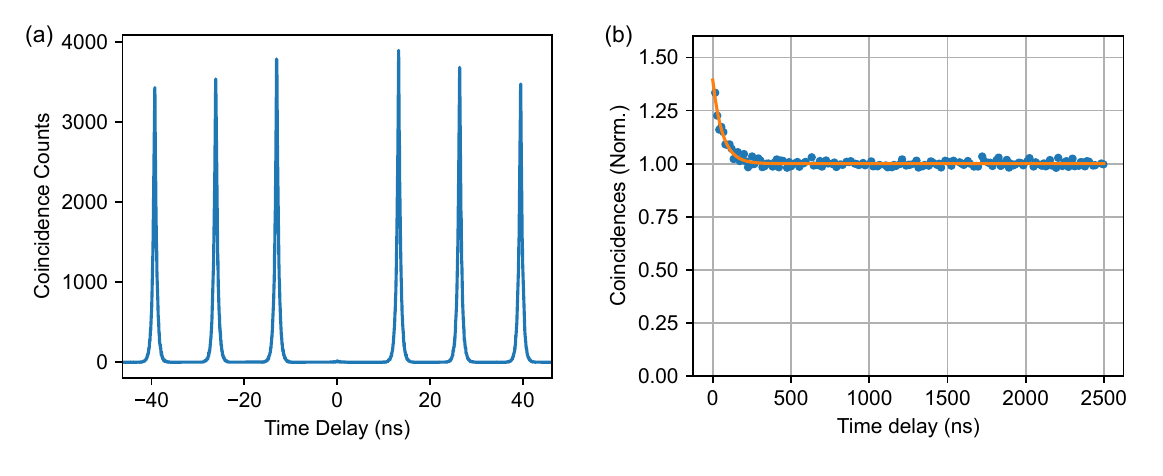}
    \caption{(a) HBT correlation histogram around $\tau=0$ of the biexciton transition under TPE at $\pi$-pulse. (b) Peak heights (blue dots) of the HBT correlation histogram up to $\tau=2.5\,\mu$s. From fitting the decay of these peak heights to the blinking function, we extracted a blinking ratio of $0.717+/-0.009$ and a correlation-time of $62+/-3\,$ns.}
    \label{si_fig:quad_gates_hbt_blinking_XX}
\end{figure}

\section{Two-Photon Interference} \label{si_sec:tpi}
\begin{figure}[h]
    \centering
    \includegraphics[width=0.65\linewidth]{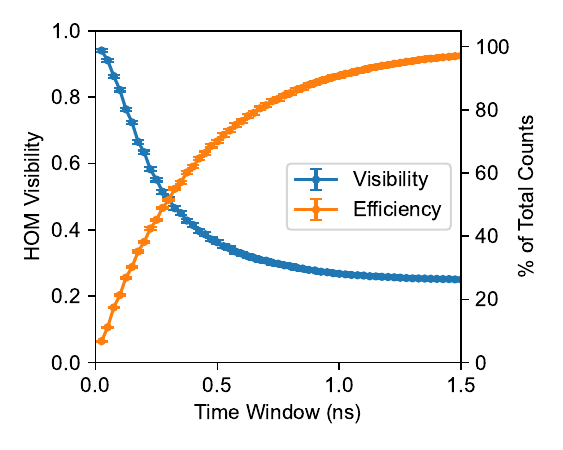}
    \caption{HOM visibility and effective source efficiency versus integration time bin $T$.}
    \label{si_fig:windowed_hom}
\end{figure}

The experimental two-photon visibility was calculated using
\begin{equation}
    V_\text{TPI}(T)  = 1 - \int_{-T}^{+T} N_{\parallel}(T)dT\Big/\int_{-T}^{+T} N_{\perp} dT   
\end{equation}
where $N_{\parallel}$ and $N_{\perp}$ are the number of coincidences in the cross-polarized and co-polarized histograms within a time window of $\pm T$.
The corresponding HOM visibility as a function of increasing time bin $T$ is plotted in SI Figure \ref{si_fig:windowed_hom}.
As expected, the TPI visibility increases as $T$ decreases with a simultaneous decrease in the source efficiency.
As $T$ increases past a few lifetimes the visibility plateaus to the quoted $24.5\%$ number calculated when $T = 5\,$ns.

\begin{figure}[!h]
    \centering
    \includegraphics[width=1\linewidth]{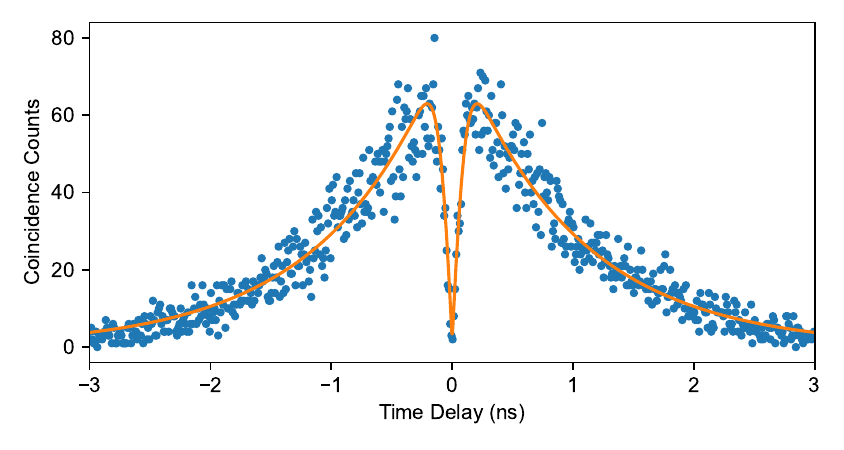}
    \caption{TPI histogram for the exciton transition under TPE with no applied bias. The total HOM visibility is $V_\text{HOM} = 8 \pm 1 \%$. The histogram is fitted to Eq. \ref{eq:kambs_HOM} to find $\tau_c = 153 \pm 2\,$ps. For the measured $\tau_X = 678$\,ps strict resonant excitation would yield a TPI visibility of $\sim 60\%$ as modeled in \cite{gangopadhyay2025mode}}
    \label{si_fig:hom_x}
\end{figure}

The function fitted to the HOM histograms is from Kambs and Becher \cite{Kambs2018-di}:
\begin{equation}
    \label{eq:kambs_HOM}
    g^{(2)}_\mathrm{HOM}(\tau) = \frac{1}{4\tau_{rad}} e^{-|\tau|/\tau_{rad}}\left(1 - e^{-2\Gamma^*|\tau| -4\pi\sigma^2\tau^2} \cos^2(\Phi)\right)
\end{equation}
where $\tau$ is the coincidence time delay, $\tau_\mathrm{rad}$ is the radiative lifetime, $\Gamma^*$ is the pure dephasing rate, $\sigma$ is the spectral diffusion rate and $\Phi$ is a parameter that encodes whether the two photons are co-polarised (indistinguishable) or cross-polarised (distinguishable).
With the values extracted by the HOM fit, the coherence length is estimated using, 
\begin{equation}
\label{eq:xc}
 \tau_c = -\frac{2\ln 2}{2\pi^2}\,\frac{\Gamma}{\varsigma^2}
 + \sqrt{\left(\frac{2\ln 2}{\pi^2}\,\frac{\Gamma}{\varsigma^2}\right)^2 + \frac{4\ln 2}{\pi^2\,\varsigma^2}}
\end{equation}
where $\varsigma = 2\sqrt{2\ln2}$ is the FWHM spectral diffusion rate \cite{Kambs2018-di}.
SI Figure \ref{si_fig:hom_x} shows a fit to the co-polarized TPI histogram of the X emission from the NWQD.

\clearpage

\section{Quadrupolar-gated NWQD Simulations}\label{si_sec:quad_gate_sim}
Numerical simulations were performed to study the effect of a lateral electric field on an InAsP quantum dot embedded in an InP nanowire. 
The goal is to predict the evolution of the excitonic wavefunction under a lateral electric field and to evaluate the resulting variation in the exciton recombination energy. 
In addition, we investigate the effect of dopants in the nanowire as a potential source of screening of the electric field experienced by the QD. 
The numerical calculations were performed using Nextnano++, a commercially available Poisson-Schrödinger solver. 
Once the geometry is defined and the material parameters are assigned, Nextnano++ solves the self-consistent Schrödinger-Poisson equations and outputs the electron and hole wavefunctions, along with the band structure and the electric field profile.

\begin{figure}[!h]
    \centering
    \includegraphics[width=1\linewidth]{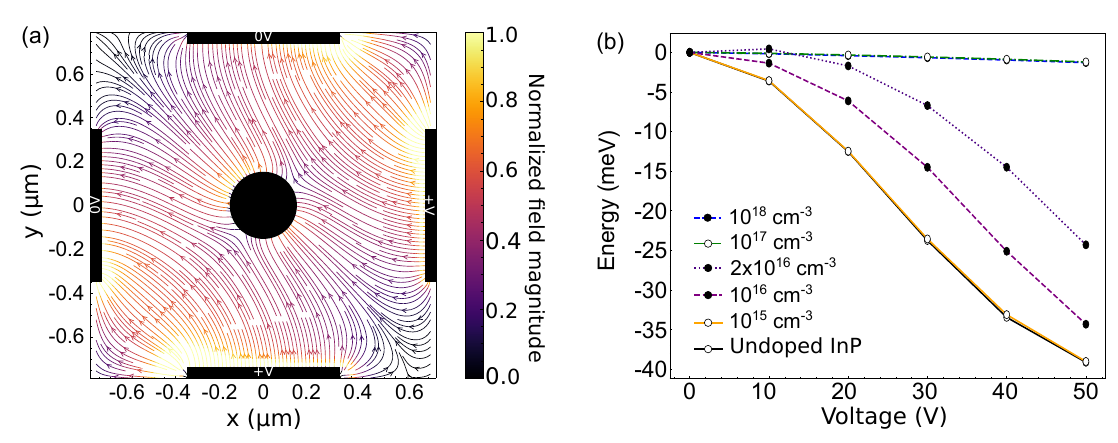}
    \caption{(a) Sample electric field distribution of the NWQD biased under a dipole configuration. (b) Effect of dopants in the InP nanowire on the tuning of the InAsP QD emission. Higher doping concentrations screen the electric field from the QD resulting in the reduction of the tuning range for the same applied bias.}
    \label{si_fig:quad_gates_simulations}
\end{figure}

For the simulation, we consider a set of quadrupole gates, each with dimensions of 700×100×50 nm (width $\times$ height $\times$ length). 
The InP nanowire is simulated with a diameter of 300 nm and a total height of 200 nm. We emphasize that the simulated height is much smaller than the actual nanowire height ($\approx12\,\mu$m). 
This truncation is necessary to restrict the simulation domain, which reduces the computational complexity and ensures a reasonable convergence time of the Poisson-Schrödinger solver. 
The QD is centred within the InP nanowire and aligned with the top of the quadrupole gates. 
The QD is assumed to have a radius of 10 nm and a height of approximately 5.5 nm, with the height determined by matching the transition energy of the simulated QD to experimental data. 
The QD composition is taken to be InAs$_{0.25}$P$_{0.75}$. 
An air gap of 600 nm is modelled between each gate and the edge of the nanowire. 
The nanowire is placed on a Si substrate, and the entire simulation is performed at a temperature of 4 K. 
A fixed voltage is applied to the gates as a boundary condition, with two neighbouring gates biased at $+V_0$ and the remaining two grounded, thereby producing a dipole configuration that generates a lateral electric field on the QD.
A sample of the dipolar field generated by the quadrupolar gates created by the Nextnano simulation is shown in SI Figure \ref{si_fig:quad_gates_simulations}a.

Unlike atomistic simulations, Nextnano++ is a single-particle solver and therefore does not explicitly model the Coulomb interaction between the electron and hole within the quantum dot. 
When an electric field is applied, the simulation captures the variation of the bandgap energy but does not account for the blueshift arising from the reduction of the electron-hole Coulomb interaction.
Experimentally, we observe a spectral shift of only a few gigahertz, which is significantly smaller than the value predicted by Nextnano++. 
While part of this discrepancy can be attributed to the absence of Coulomb interactions in the simulation, it may also arise from the assumption of a dopant-free nanowire. 
In Nextnano++, materials can be doped either n-type or p-type. 
To investigate this effect, the dopant concentration was swept from $10^{15}$ to $10^{19}$ cm$^{-3}$.
We plot the estimated Stark shift in Figure \ref{si_fig:quad_gates_simulations}b for these various doping concentrations.
We observe a strong reduction of the wavelength shift at higher doping levels.
Indeed, the presence of free carriers screens the applied lateral electric field, thereby reducing the effective field experienced by the quantum dot.

\end{document}